\documentclass[
  ,draft            % use draft while you are working on the paper
  ]
  {aipproc}
\pdfoutput=1
\layoutstyle{6x9}

\begin{document}

\title{The Coexistence of Classical Bulges, Pseudobulges, and Supermassive Black Holes}

%\classification{<Replace this text with PACS numbers; choose from this list:
%                \texttt{http://www.aip..org/pacs/index.html}>}

\author{Peter Erwin}{
  address={Max-Planck-Institut f\"{u}r extraterrestrische Physik,
Giessenbachstrasse,
D-85748 Garching, Germany}
  ,altaddress={Universit\"{a}ts-Sternwarte M\"{u}nchen,
Scheinerstrasse 1,
D-81679 M\"{u}nchen, Germany}
}

%\author{<author2>}{
%  address={<common address for author2 and author3>}
%}
%
%\author{<author3>}{
%  address={<common address for author2 and author3>}
%  ,altaddress={<author1 address>} % additional visiting address
%}

\begin{abstract}
Some S0 and early-type spiral galaxies possess ``composite bulges''; in these
galaxies, the photometric bulge -- the central stellar
light in excess of the disk light -- is composed of both a ``(disky)
pseudobulge'', with a flattened, disklike morphology and relatively cool
stellar kinematics, and a rounder, kinematically hot ``classical bulge''
embedded within.  I speculate that supermassive black holes in such
galaxies may correlate with the classical-bulge component only, and not
with the pseudobulge component; preliminary comparisons with SMBH masses
appear to support this hypothesis.
\end{abstract}

\maketitle

%%%%%%%%%%%%%%%%%%%%%%%%%%%%%%%%%%%%%%%%%%%%
%% MAINMATTER
%%%%%%%%%%%%%%%%%%%%%%%%%%%%%%%%%%%%%%%%%%%%

\section{Introduction}

There are now several well-established correlations between central
supermassive black holes (SMBHs) and their host galaxies.  These take
the form of correlations between the SMBH mass and properties of the
bulge, such as its central stellar velocity dispersion, luminosity, or
total mass, and are thought to reflect strong ties -- perhaps even
identities -- between the mechanisms that fuel SMBH growth (and
accompanying nuclear activity) and the buildup of bulges.  The term
``bulge'' is usually taken to mean a kinematically hot stellar spheroid. 
In the case of elliptical galaxies, this is the galaxy itself; in the
case of disk galaxies (S0 and spirals), this is assumed to be the
central ``photometric excess'' -- the excess stellar population above
that of the galaxy disk.

If all disk-galaxy bulges were the same kind of structure as elliptical
galaxies -- which is in fact the traditional view of bulges -- then
things would be relatively simple: theorists could argue for mechanisms
which fuel SMBHs and grow bulges (e.g., via rapid, violent mergers and
accompanying starbursts) across the Hubble sequence.  But recent
observations and arguments suggest that the ``bulges'' in many disk
galaxies may be a different kind of beast: a disklike structure which
has developed slowly (``secularly'') out of the disk and retains many
disk properties \citep[see the review by][]{kk04}.  If SMBHs correlate
with such ``pseudobulges'' to the same degree as they do with classical
bulges and ellipticals, then explaining these correlations becomes much
harder, because the respective formation mechanisms may be very
different.

Here, I discuss the idea that some galaxies may have ``composite
bulges,'' with both classical bulges \textit{and} pseudobulges
\citep[see also][]{erwin03,athanassoula05}. I conjecture that SMBHs in
such galaxies may correlate with the classical bulge component alone,
rather than with the pseudobulge, thus preserving the SMBH-bulge
correlations as correlations between SMBHs and kinematically hot
spheroids. Observations are currently in progress to test this
hypothesis.

\section{Definitions and an Example}

The term \textbf{photometric bulge} means the central excess stellar
light of the galaxy, above that of the (extrapolated) exponential disk,
regardless of morphology or kinematics.  The photometric bulge can be
identified by standard bulge-disk decompositions, and is in fact what
most people doing bulge-disk decompositions mean by ``bulge.''  If some
or all of the photometric bulge turns out to be morphologically disklike
-- flattening similar to that of the outer disk, possibly containing
disk features such as nuclear bars and rings -- and kinematically cool
(e.g., local stellar $V/\sigma > 1$), then this a \textbf{(disky)
pseudobulge}. (``Disky'' is used to avoid confusion with other
structures which have been called pseudobulges, such as the vertically
thickened ``boxy/peanut-shaped bulges'' produced by bars.)  Finally, a
central component which is clearly rounder than the outer disk and which
is kinematically hot is a \textbf{classical bulge}, under the hypothesis
that this is closer to what has been traditionally meant by the term
``bulge.''

\begin{figure}
  \includegraphics[height=.3\textheight]{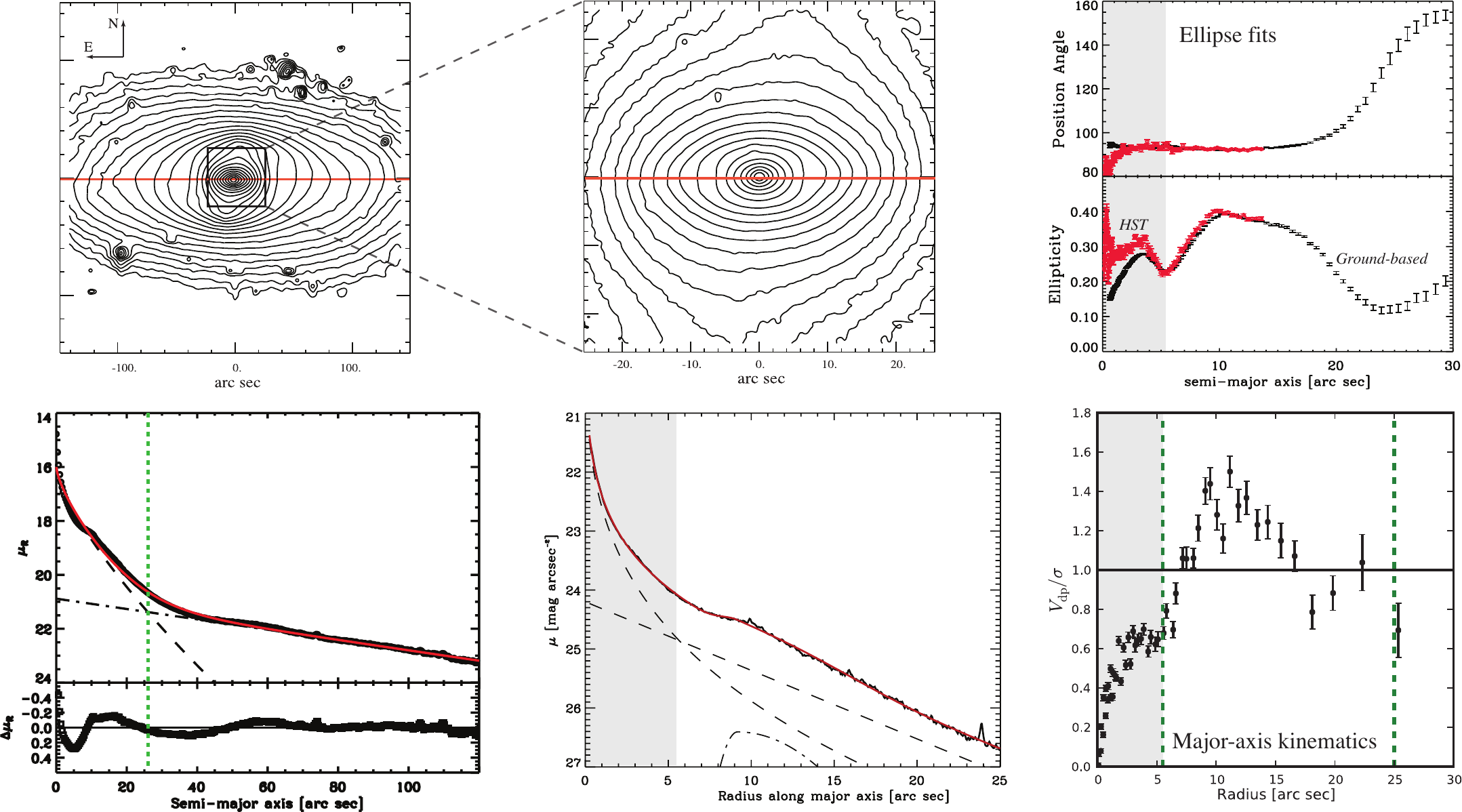}
  \caption{The composite bulge of NGC 4371. Upper left: $R$-band
  isophotes, showing the outer disk, the bar, and photometric bulge; red
  line indicates major axis. Lower left: major-axis surface brightness
  profile with bulge-disk decomposition; vertical dashed green line
  indicates boundary of photometric bulge. Upper middle: inner
  isophotes, showing disky pseudobulge region.  Lower middle: surface
  brightness profile with bulge/disk/nuclear-ring decomposition
  (\textit{not} the same as the first decomposition); gray zone
  indicates the classical bulge region. Upper right: ellipse fits to
  isophotes.  Lower right: $V_{\rm dp} / \sigma$ as a function of radius
  (see text), based on folded major-axis long-slit spectra.}\label{fig:n4371}
\end{figure}

NGC~4371 is a barred S0 in the Virgo Cluster.  The left-hand
panels of Fig.~\ref{fig:n4371} show the major-axis surface-brightness
profile (based on ground-based and \textit{HST} optical images).
Superimposed is a ``naive'' bulge/disk decomposition, where I assume
that the light is the sum of an outer exponential and an inner S\'ersic
component.  The photometric bulge is then identified as the region where
the S\'ersic component dominates ($r < 30^{\prime\prime}$).

Closer examination shows that part of the photometric bulge region is
quite elliptical, and harbors a bright nuclear ring with radius
$\sim10^{\prime\prime}$ \citep{erwin99}. \citet{kormendy82} used the
ratio of the peak stellar velocity in this region to the central
velocity dispersion ($V_{\rm max}/\sigma_{0}$) to argue that the
photometric bulge was rotating faster than an isotropic oblate
rotator (a simple model for classical bulges) would.  The lower-right
panel of Fig.~\ref{fig:n4371} shows $V_{\rm dp} / \sigma$ as a function
of radius, where $V_{\rm dp}$ is the stellar velocity deprojected to its
in-plane value and $\sigma$ is the stellar velocity dispersion.  This
ratio rises to $\sim1.5$ in the vicinity of the nuclear ring,
indicating that the stellar kinematics are (relatively) cool and
disklike.

As the residuals in the lower-left panel of Fig.~\ref{fig:n4371}
indicate, the surface-brightness profile in the photometric bulge region
is not actually a simple S\'ersic profile.  We can model the inner $\sim
30^{\prime\prime}$ as the sum of an exponential and a separate, inner
S\'ersic component (along with a small contribution from the nuclear
ring; lower middle panel of Fig.~\ref{fig:n4371}).  When we do this, we
see that this inner S\'ersic component matches quite well with a region
of \textit{rounder} isophotes, at $r < 5^{\prime\prime}$. Moreover,
$V_{\rm dp} / \sigma$ has a plateau of $\sim 0.7$ in this region,
suggesting that we are seeing a separate, kinematically hotter
component.  This, then, is evidence for a classical bulge embedded
within the disky pseudobulge.

%%%%%%%%%%%%%%%%%%%%%%%%%%%%%%%%%%%%%%%%%%%%
%% Sample figure:
%%
%% The option [height=...] scales the picture to the given height,
%% without it it would be printed at its nominal size
%%%%%%%%%%%%%%%%%%%%%%%%%%%%%%%%%%%%%%%%%%%%

\section{Results and Discussion}

A total of ten composite-bulge systems have been identified and analyzed
in at least preliminary fashion; all but two of these are S0 galaxies
(the others are early-type spirals).  Although the collection assembled
so far is \textit{not} based on any rigorous, unbiased sample,
comparison with a partially complete survey of local S0 galaxies
suggests that at least $\sim 20$\% of S0 galaxies may have composite
bulges.

In these systems, the majority of the stellar light making up the
photometric bulge is in the disky pseudobulge; the classical bulge is
typically only $\sim 25$\% as bright as the pseudobulge component.
Compared to the whole galaxy, the classical bulges have a median B/T of
only 0.1, with a range of 0.02--0.22. They have S\'ersic indices ranging
from 1.3 to 3.4, with a median value of 2.0, and are quite compact:
half-light radii range from 70 pc to 1000 pc (median = 160 pc). Note
that even the smallest of these classical bulges is more than an order
of magnitude larger than a typical nuclear star cluster \citep[$R_e =
2$--5 pc;][]{boker08}, and there is at least one clear case of a nuclear
star cluster existing as a distinct component \textit{inside} the
classical bulge.

\begin{figure}
  \includegraphics[height=.3\textheight]{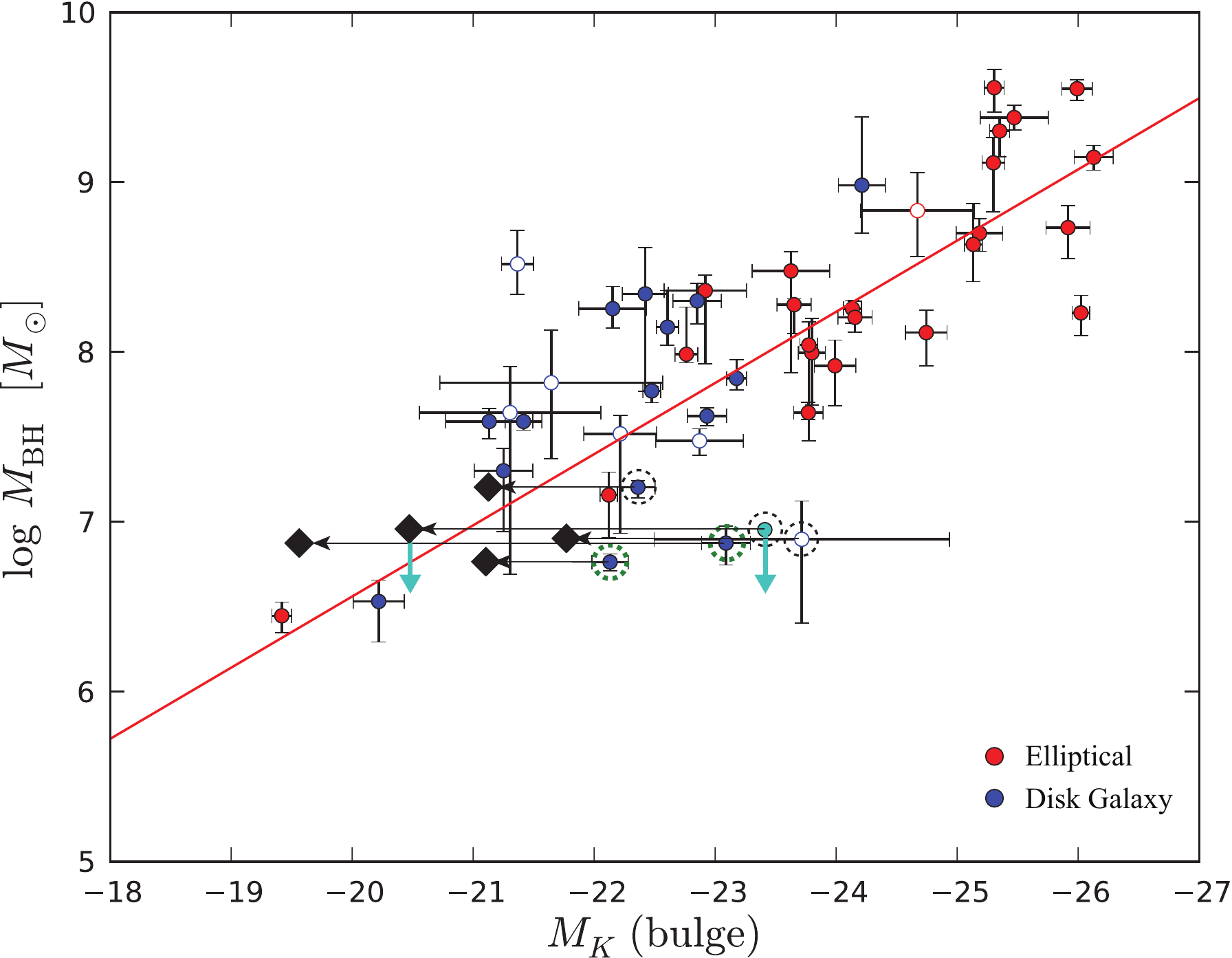}
  \caption{The black hole--bulge relation (SMBH mass versus $K$-band
  luminosity of the photometric bulge), based on
  \citet{erwin-gadotti09}, with additional data from \citet{nowak09}. 
  The five circled points are composite-bulge galaxies; the horizontal
  black arrows link the luminosity of the entire photometric bulge
  (right) to the luminosity of the embedded classical-bulge component
  only (left).  Note that the latter are a better match to the overall
  correlation (diagonal red line).  Blue downward arrows indicate
  upper limit on SMBH mass for NGC 3945 from
  \citet{gultekin09}.}\label{fig:relation}
\end{figure}

The coexistence of classical bulges and pseudobulges suggests a
potentially interesting hypothesis concerning the relationship between
central supermassive black holes and their parent galaxies.  If SMBHs
correlate (and are formed in concert with) \textit{photometric} bulges,
then they should correlate with pseudobulges in pseudobulge-dominated
cases.  (Which in turn implies that pseudobulge and SMBH growth must be
linked in a fashion not too dissimilar from that linking SMBH mass and
elliptical galaxies, despite the radically different scenarios for
pseudobulge and elliptical galaxy growth.)  On the other hand, if the
true correlation is between SMBHs and kinematically hot spheroids, then
we should expect the strongest correlation to be with the embedded
classical bulge rather than with the photometric bulge.  It is unclear
how one would separate out classical and pseudobulge contributions to
the central velocity dispersion; but separating out their respective
contributions to bulge \textit{luminosity} (or stellar mass) is
relatively simple. Fig.~\ref{fig:relation} is a preliminary exploration
of this question: do SMBH masses correlate with total photometric bulge
luminosities, or only with the classical bulge components?  The figure
includes recent SMBH measurements with VLT-SINFONI for two of the
composite-bulge systems \citep{nowak09}; further observations of other
candidates are being analyzed.

%%%%%%%%%%%%%%%%%%%%%%%%%%%%%%%%%%%%%%%%%%%%%%%%
%% BACKMATTER
%%%%%%%%%%%%%%%%%%%%%%%%%%%%%%%%%%%%%%%%%%%%%%%%

\begin{theacknowledgments}
I would like to thank my collaborators in the various projects this
research is part of and derived from, including Juan Carlos Vega
Beltr\'an, John E. Beckman, Dimitri Gadotti, Roberto Saglia, Nina Nowak,
Jens Thomas, and Ralf Bender.  This work was supported by Priority
Programme 1177 of the Deutsche Forschungsgemeinschaft.
\end{theacknowledgments}



\begin{thebibliography}{9}

\bibitem[Athanassoula(2005)]{athanassoula05} E. Athanassoula, \emph{MNRAS}
 \textbf{358}, 1477--1488 (2005).
 
\bibitem[B\"oker(2008)]{boker08} T. B\"oker, in \emph{The Universe under
the Microscope -- Astrophysics at High Angular Resolution}, edited by
R. Schoedel et al., \emph{Journal of Physics: Conference Series}
\textbf{131}, pp. 012043 (2008).

\bibitem[Erwin \& Sparke(1999)]{erwin99} P. Erwin and L. S. Sparke, \emph{ApJL} 
\textbf{521}, L37--40 (1999).

\bibitem[Erwin \& Gadotti(2010)]{erwin-gadotti09}
P. Erwin and D. Gadotti, in prep. (2010).

\bibitem[Erwin et al.(2003)]{erwin03} P. Erwin, J. C. Vega Beltr\'an, A. W.
Graham, and J. E. Beckman, \emph{ApJ} \textbf{597}, 929--947 (2003).

\bibitem[G\"ultekin(2009)]{gultekin09} K. G\"ultekin et al., \emph{ApJ} \textbf{695},
1577--1590 (2009).

\bibitem[Kormendy(1982)]{kormendy82} J. Kormendy, \emph{ApJ} \textbf{257}, 75--88 (1982).

\bibitem[Kormendy \& Kennicutt(2004)]{kk04} J. Kormendy and R. C.
Kennicutt, Jr., \emph{Ann. Rev. Astron. Astrophys.} \textbf{42},
603--683 (2004).

\bibitem[Nowak et al.(2009)]{nowak09}
N. Nowak, J. Thomas, P. Erwin, R.~P. Saglia, R. Bender, and R.~I. Davies, 
\emph{MNRAS}, submitted (2009).


\end{thebibliography}
\end{document}